\documentclass[twocolumn,superscriptaddress,letter]{revtex4}%
\usepackage{amssymb}
\usepackage{color}
\usepackage{graphicx}
\usepackage{dcolumn}
\usepackage{bm}
\usepackage[header,title,page,titletoc]{appendix}
\usepackage{amsmath}
\usepackage{subfigure}
\usepackage{amsfonts}%
\providecommand{\U}[1]{\protect \rule{.1in}{.1in}}

\begin{document}
\title{Theory of Vacancy-Induced Intrinsic Magnetic Impurity with Quasi-Localized
Spin Moment in Graphene}
\author{Yang Li}
\thanks{These authors contributed equally to this paper.}
\affiliation{Department of Physics, Beijing Normal University, Beijing, 100875, P. R. China}
\author{Jing He}
\thanks{These authors contributed equally to this paper.}
\affiliation{Department of Physics, Hebei Normal University, HeBei, 050024, P. R. China}
\author{Su-Peng Kou}
\thanks{Corresponding author}
\email{spkou@bnu.edu.cn}
\affiliation{Department of Physics, Beijing Normal University, Beijing, 100875, P. R. China}

\begin{abstract}
In this paper, by considering the Hubbard model on a honeycomb lattice, we
developed a theory for the intrinsic magnetic impurities (MIs) with the
quasi-localized spin moments induced by the vacancies in graphene. Because the
intrinsic MIs are characterized by the zero modes that {are orthotropic to the
itinerant electrons, }their properties are much different from those of
Anderson MIs with the well-localized spin moments.

\end{abstract}
\maketitle

Graphene consists of carbon atoms organized into honeycomb lattice, coupling
with each other through $\mathrm{sp}^{2}$ orbitals. Since it was isolated in
2004 by A. Geim and K. Novoselov\cite{nov1,nov2}, it had been intensively
studied in the last several years. The fast uptake of interest in graphene is
due primarily to its exceptional properties\cite{neto,sarma}. The magnetic
impurities in the graphene becomes an interesting issue due to the possible
application to the spintronics\cite{aw}. In particular, people found that the
missing atom (single vacancy) in graphene may induce a quasi-localized state
(the so-called zero mode) distributed around the
impurity\cite{zero-mode,ya,zero1}. The induced spin moment by the
lattice-defects was also observed in experiments\cite{nair12,hong12}. Based on
the Anderson model or the Kondo model, the Kondo effect and the RKKY
{interaction between magnetic impurities have }studied by variety of
groups\cite{rkky,kondo,kondo1,fuhrer11}. People found that due to the linear
dispersion in graphene, some features of the magnetic impurities are changed.
For example, at half filling, the RKKY coupling is strictly ferromagnetic (FM)
for spin moments on the same sublattice and antiferromagnetic (AFM) for spin
moments on different sublattices, in both cases falling off as $1/R^{3}%
$\cite{rkky}. The $1/R^{3}$ decay rate differs from the usual $1/R^{2}$ decay
rate for the magnetic impurities in two dimensional Fermi liquid\cite{rkky}.

In this paper we study the properties of the vacancy-induced magnetic
impurities of the graphene by considering the Hubbard model on a honeycomb
lattice. Here, the magnetic impurities are induced by removing an atom rather
than doping an extra magnetic impurity. The wave function of zero modes around
the magnetic impurity is known to be eigenstates of the system and
\emph{orthogonal} to the itinerant electronic states. Thus, the Anderson model
or its deduced Kondo model is \emph{not} applicable. On the other hand, the
particle density of the zero modes falls off as $1/R^{2}$\cite{zero-mode}. The
absence of a localized length indicates a \emph{quasi-localized} spin moment
(QLSM) rather than a well-localized spin moment (WLSM). And the QLSM will
never be screened by the itinerant electrons. As a result, this type of
magnetic impurities is much difference from the traditional Anderson
impurities, which the itinerant electrons always try to screen. To emphasize
the differences, we call it "\emph{intrinsic magnetic impurity}" to
distinguish the traditional Anderson impurity which can be classified to
"\emph{extrinsic magnetic impurity}". So one can imagine that the system turns
to repel the extrinsic magnetic impurities (MIs) by screening them, while
accepting the intrinsic MIs. In this paper, our task is to systematically
recognize the properties of the intrinsic MIs on graphene.

Our starting point is the Hubbard model on a honeycomb lattice, of which the
Hamiltonian is
\begin{equation}
H=-t\sum \limits_{\left \langle {i,j}\right \rangle ,\sigma}\hat{c}_{i\sigma
}^{\dagger}\hat{c}_{j\sigma}+U\sum_{i}n_{i\uparrow}n_{i\downarrow}-\mu
\sum \limits_{{i}}\hat{c}_{i}^{\dagger}\hat{c}_{i}%
\end{equation}
where $t$ is the nearest neighbor hopping, $\mu$ is the chemical potential and
$U$ is the strength of the repulsive interaction, respectively. For graphene,
$t$ is about $2.8\mathrm{eV}$, $U$ is about $1.5t\simeq4.2\mathrm{eV}$. In
this paper we ignore the next nearest neighbor hopping.

Since the honeycomb lattice is a bipartite lattice, we have two sublattices,
\textrm{A} sublattice and \textrm{B} sublattice. In momentum space, for free
electrons, $H$ is reduced into $H_{\mathrm{Free}}=\sum_{\mathbf{k}}%
[\epsilon(\mathbf{k})\hat{c}_{\mathbf{k}\mathrm{A}}^{\dagger}\hat
{c}_{\mathbf{k}\mathrm{B}}+\epsilon^{\ast}(\mathbf{k})\hat{c}_{\mathbf{k}%
\mathrm{B}}^{\dagger}\hat{c}_{\mathbf{k}\mathrm{A}}]$ where $\epsilon
(\mathbf{k})=\sum_{\mathbf{\delta}}-te^{i\mathbf{k}\cdot \mathbf{\delta}},$ in
which $\mathbf{k}=(k_{x},k_{y})$ is momentum in reduced Brillouin Zone (BZ),
$\mathbf{\delta}$ are nearest neighbor links. After diagnalization of the
Hamiltonian, the spectra become $E_{\pm}\left(  \mathbf{k}\right)
=\pm \left \vert \epsilon(\mathbf{k})\right \vert $. Near the nodal points
$\mathbf{K}_{1}=(0,\frac{4\pi}{3\sqrt{3}}),$ $\mathbf{K}_{2}=(\frac{2\pi}%
{3},\frac{2\pi}{3\sqrt{3}}),$ the dispersion becomes a linear one as $E_{\pm
}\left(  k\right)  \simeq v_{F}\left \vert \mathbf{k}\right \vert $ where
$v_{F}$ is the Fermi velocity of the electrons. The lattice constant is set to
be unit in the following calculations.

Firstly, we study the graphene with a lattice defect on \textrm{A} sublattice
in the presence of an on-site potential at site $\mathbf{R}$, $H\rightarrow
H(U=0)+V_{\mathbf{R}}\hat{c}_{\mathbf{R}}^{\dagger}\hat{c}_{\mathbf{R}}.$ In
the unitary limit, the lattice defect becomes a vacancy, of which we have an
infinite on-site potential, i.e., $V_{\mathbf{R}}\rightarrow \infty$. Two
localized states that are orthogonal to the itinerant electronic states
appear, one for the electrons with up spin, and the other for the electrons
with down spin. Due to the particle-hole symmetry, the localized state around
the vacancies have exactly zero energy and its wave-function distributes only
on \textrm{B} sublattice\cite{comment}. In the continuum limit, the wave
function of the zero mode introduced by one vacancy has the form of $\psi
_{0}\left(  \mathbf{r}\right)  =\frac{e^{i\mathbf{K}_{1}\cdot \mathbf{r}}%
}{x+iy}+\frac{e^{i\mathbf{K}_{2}\cdot \mathbf{r}}}{x-iy}$, $\mathbf{r}%
=(x,y)$\cite{zero-mode}. Far from the vacancy, the decay rate of the particle
density is $\left \vert \psi_{0}\left(  \mathbf{r}\right)  \right \vert
^{2}\rightarrow1/\left \vert \mathbf{r}\right \vert ^{2}$. So we call it
quasi-localized state.

When we consider the on-site interaction, there exists effective repulsive
interaction between the electrons trapped on the zero mode as $U_{\mathrm{eff}%
}\hat{n}_{\mathbf{R}\uparrow}\hat{n}_{\mathbf{R}\downarrow}$ where
$U_{\mathrm{eff}}=U\sum_{i}|\psi_{0}(i)|^{4}.$ $\hat{n}_{\mathbf{R}}$ is the
number operator of quasi-localized state. After considering the chemical
potential term, we get the effective Hamiltonian of the electrons on the zero
mode around a single vacancy $\hat{H}_{\mathrm{L}}=U_{\mathrm{eff}}\hat
{n}_{\mathbf{R}\uparrow}\hat{n}_{\mathbf{R}\downarrow}-\mu_{\mathrm{eff}}%
(\hat{n}_{\mathbf{R}\uparrow}+\hat{n}_{\mathbf{R}\downarrow})$ where
$\mu_{\mathrm{eff}}=\mu$. For the case of $U_{\mathrm{eff}}<\mu_{\mathrm{eff}%
},$ the zero mode is double occupied; For the case of $\mu_{\mathrm{eff}}<0,$
the zero mode is empty. So the spin moment of the quasi-localized state (we
call it quasi-localized spin moment) exists when a finite chemical potential
is smaller than $U_{\mathrm{eff}}$ as $U_{\mathrm{eff}}>\mu_{\mathrm{eff}}>0.$

Because the wave-functions of the quasi-local states (we borrow the name
"d-orbitals" to label them) $\left \vert d\right \rangle $ and those of the
itinerant electrons (we borrow the name "s-orbitals" to label them)
$\left \vert s,\mathbf{k}\right \rangle $ are always orthogonal each other,
$\left \langle s,\mathbf{k}\mid d\right \rangle =0,$ there is no "s-d
hybridization" between the zero modes and those of the itinerant electrons.
Instead, when we consider the on-site particle interaction, there exists
effective "s-d coupling" between the QLSM induced by the vacancy and the spin
moments of the itinerant electrons. Such s-d coupling between the QLSM and the
itinerant electrons can be regarded as the Hund rule's coupling for two
orthogonal orbitals - an orbital of zero mode and an orbital with finite
wave-vectors. As a result, the s-d coupling is always ferromagnetic and
momentum-dependence. All these features are universal for a vacancy-induced MI
in graphene, a remarkable example of the intrinsic MIs.

\begin{figure}[ptb]
\includegraphics* [width=0.5\textwidth]{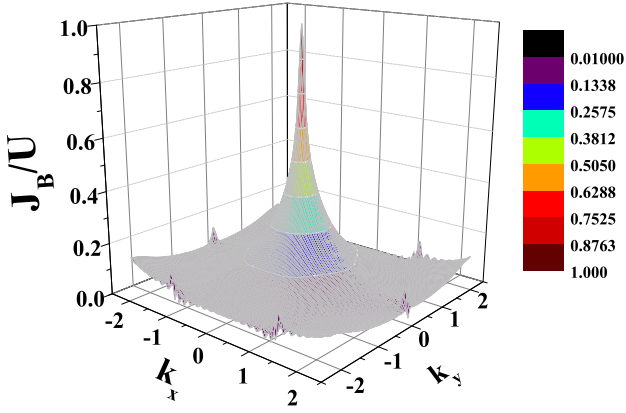}\caption{(Color online) The
s-d coupling $J_{\mathrm{B}}(\mathbf{q})$ in momentum space. At $\mathbf{k}%
=(0,0),$ $J_{\mathrm{B}}(\mathbf{q})=U$.}%
\end{figure}

We assume there exists QLSM on the vacancy (or $U_{\mathrm{eff}}%
>\mu_{\mathrm{eff}}>0$). The s-d coupling which describes the process that the
itinerant electrons are scattered by the quasi-localized state from
$\mathbf{k}$ to $\mathbf{k}^{\prime}$ is given by the following non-local FM
Kondo-like Hamiltonian
\begin{align}
H_{\mathrm{s-d}}  &  =-\sum_{\mathbf{k},\mathbf{k}^{\prime}}J_{\mathrm{A}%
}\left(  \mathbf{k}^{\prime}-\mathbf{k}\right)  \mathbf{\hat{S}}_{\mathbf{R}%
}\cdot \mathbf{\hat{s}}_{\mathrm{A},\mathbf{kk}^{\prime}}\nonumber \\
&  -\sum_{\mathbf{k},\mathbf{k}^{\prime}}J_{\mathrm{B}}\left(  \mathbf{k}%
^{\prime}-\mathbf{k}\right)  \mathbf{\hat{S}}_{\mathbf{R}}\cdot \mathbf{\hat
{s}}_{\mathrm{B},\mathbf{kk}^{\prime}}%
\end{align}
where $\mathbf{\hat{S}}_{\mathbf{R}}$ is the spin operator of the QLSM induced
by the vacancy at site $\mathbf{R}$ and $\mathbf{\hat{s}}_{\mathbf{kk}%
^{\prime}}$ is the spin operator of the itinerant electrons on \textrm{A}%
/\textrm{B} sublattice $\mathbf{\hat{s}}_{\mathrm{A}/\mathrm{B},\mathbf{kk}%
^{\prime}}=c_{\mathrm{A}/\mathrm{B,}\mathbf{k}}^{\dagger}\mathbf{\sigma
}c_{\mathrm{A}/\mathrm{B,}\mathbf{k}^{\prime}}.$ $J_{\mathrm{A}}%
(\mathbf{k}^{\prime}-\mathbf{k})$ ($J_{\mathrm{B}}(\mathbf{k}^{\prime
}-\mathbf{k})$) is the strength of the s-d coupling on sublattice $\mathrm{A}$
($\mathrm{B}$), respectively as
\begin{align}
J_{\mathrm{A/B}}(\mathbf{k}^{\prime}-\mathbf{k})  &  =\int \int \psi
_{\mathbf{k},\mathrm{A/B}}^{\ast}\left(  \mathbf{r}_{1}\right)  \psi_{0}%
^{\ast}\left(  \mathbf{r}_{2}\mathbf{-R}\right) \nonumber \\
&  \times U\left(  \mathbf{r}_{1}\mathbf{-r}_{2}\right)  \psi_{\mathbf{k}%
^{\prime},\mathrm{A/B}}\left(  \mathbf{r}_{2}\right)  \psi_{0}\left(
\mathbf{r}_{1}\mathbf{-R}\right)  d^{2}\mathbf{r}_{1}d^{2}\mathbf{r}%
_{2}\nonumber \\
&  =\frac{U}{N}\sum_{j\in \mathrm{A/B}}\left \vert \psi_{0,j}\right \vert
^{2}e^{i(\mathbf{k}^{\prime}-\mathbf{k})\cdot \mathbf{R}_{j}}%
\end{align}
where the interaction $U\left(  \mathbf{r}_{1}\mathbf{-r}_{2}\right)
=U\delta \left(  \mathbf{r}_{1}\mathbf{-r}_{2}\right)  $. $\psi_{\mathbf{k}%
\mathrm{A}}\left(  \mathbf{r}_{1}\right)  $ is the field of itinerant
electrons with wave vector $k$. We see that the s-d coupling is induced by the
on-site interaction $U$, and it is non-local. So, we call it "\emph{nonlocal
s-d coupling}" of QLSM.

Because the quasi-localized state around the vacancy at \textrm{A} sublattice
distributes only on \textrm{B} sublattice, we have $J_{\mathrm{A}}%
(\mathbf{k}^{\prime}-\mathbf{k})=0,$ $J_{\mathrm{B}}(\mathbf{k}^{\prime
}-\mathbf{k})\neq0$. In Fig.(1) we show the non-local s-d coupling in the
momentum space. Form Fig.(1), one can see that the non-local s-d coupling has
a maximum value at $\mathbf{q}=\mathbf{k}^{\prime}-\mathbf{k}\rightarrow0$,
$\left \vert J_{\mathrm{B}}(\mathbf{q}=\mathbf{0})\right \vert =U$ and then
falls off as $\left \vert J_{\mathrm{B}}(\mathbf{q})\right \vert \rightarrow
1/\left \vert \mathbf{q}\right \vert .$ The situation of QLSM is much different
from the WLSM of Anderson magnetic impurity due to the s-d hybridization, of
which the local AFM s-d coupling (the Kondo coupling) is constant in momentum
space and can be written into a formulation into real space, $H_{\mathrm{s-d}%
}=J\mathbf{\hat{S}}_{i}\cdot \mathbf{\hat{s}}_{i}$.

Now we have a nonlocal FM Kondo model that describes coupling between the QLSM
to the itinerant electrons due to the Hund rules' coupling $H_{\mathrm{s-d}}.$
The Hamiltonian becomes
\begin{equation}
H_{\mathrm{kondo}}=H_{\mathrm{Free}}-\mu \sum \limits_{\mathbf{k}}\hat
{c}_{\mathbf{k}}^{\dagger}\hat{c}_{\mathbf{k}}+H_{\mathrm{s-d}}.
\end{equation}
We have used a mean-field approach to study the possible Kondo effect. From
the mean-field theory, we don't find the bound state between the QLSM and the
itinerant electrons. Thus we \emph{guess} that the quasi-local state has 1/2
spin moment and always decouple from the itinerant electrons.

\begin{figure}[ptb]
\includegraphics* [width=0.5\textwidth]{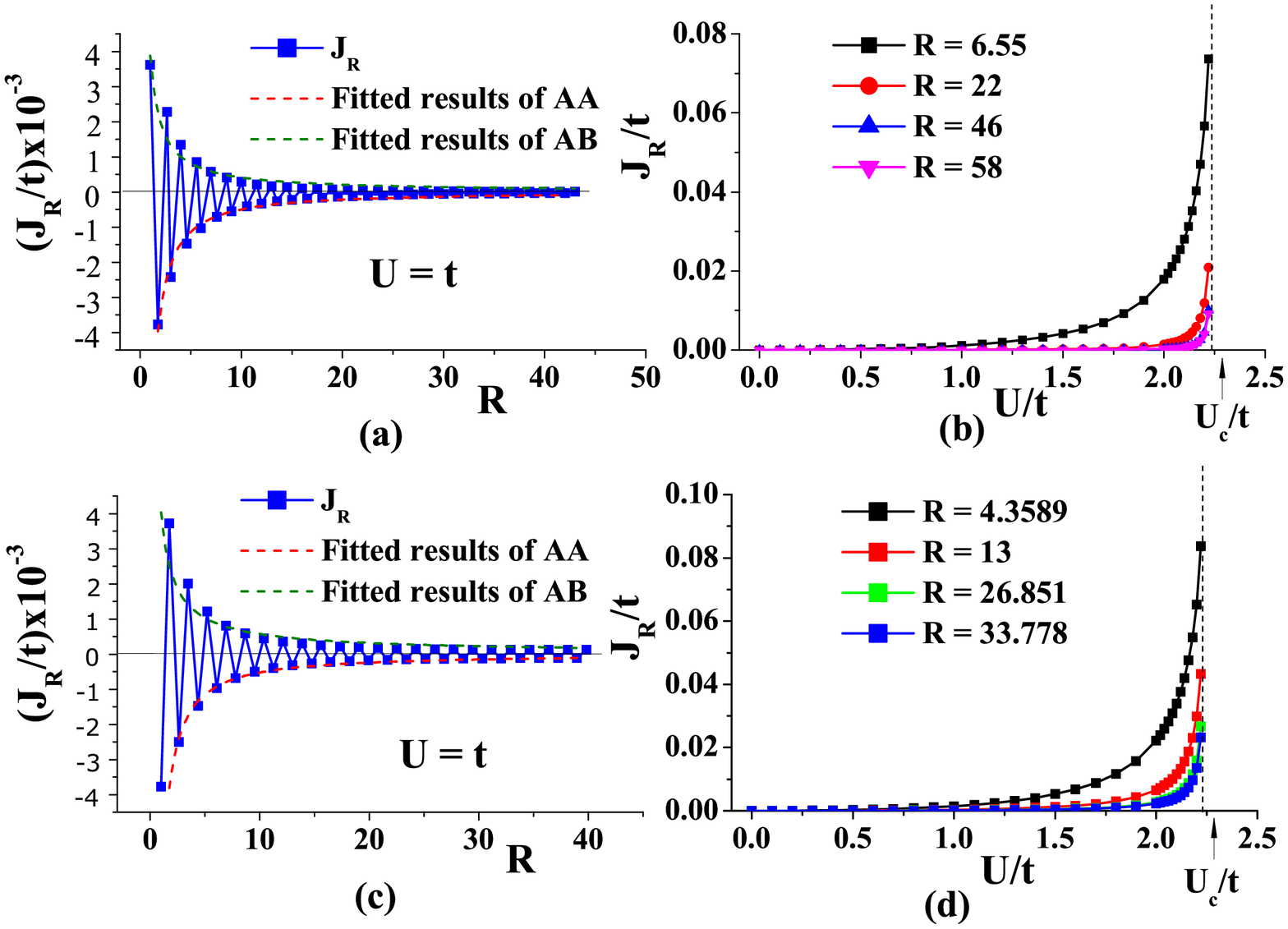}\caption{(Color online) The
RKKY coupling between two QLSMs. For (a) and (c), the interaction strength is
$U=t$. In (b) and (d), $J_{R}$ diverges at $U_{c}=2.23t.$ (a) and (b) are the
results for armchair direction. (c) and (d) are the results for zigzag
direction. The fitted line is $1/R$.}%
\end{figure}

In the following parts, we consider the case of two vacancies. Due to
quasi-localization, the wave functions of the zero modes around two vacancies
could overlap even when they are not close to each other. The overlap of the
wave functions leads to the direct Heisenberg exchange coupling and the
superexchange coupling between the QLSMs. When the distance is too far to
overlap, the coupling between QLSMs mainly comes from the RKKY\ interaction
which is mediated by the itinerant electron.

At first step, we study the RKKY coupling between two QLSMs around the
vacancies on $\mathbf{R}$ and $\mathbf{R}^{\prime}$ which is described by the
following Hamiltonian
\begin{equation}
J_{\mathrm{R}}(\mathbf{R},\mathbf{R}^{\prime})\mathbf{\hat{S}}_{\mathbf{R}%
}\cdot \mathbf{\hat{S}}_{\mathbf{R}^{\prime}}%
\end{equation}
where the RKKY interaction strength $J_{\mathrm{R}}(\mathbf{R},\mathbf{R}%
^{\prime})$ is
\begin{align}
J_{\mathrm{R}}(\mathbf{R},\mathbf{R}^{\prime})  &  =-\sum_{\mathbf{q}%
}J_{\mathrm{A}\text{,}\mathbf{R}}\left(  \mathbf{q}\right)  J_{\mathrm{A}%
,\mathbf{R}^{\prime}}(-\mathbf{q})\chi_{\mathrm{AA}}(\mathbf{q})\nonumber \\
&  -\sum_{\mathbf{q}}J_{\mathrm{A}\text{,}\mathbf{R}}\left(  \mathbf{q}%
\right)  J_{\mathrm{B},\mathbf{R}^{\prime}}(-\mathbf{q})\chi_{\mathrm{AB}%
}(\mathbf{q})\nonumber \\
&  -\sum_{\mathbf{q}}J_{\mathrm{B}\text{,}\mathbf{R}}\left(  \mathbf{q}%
\right)  J_{\mathrm{A},\mathbf{R}^{\prime}}(-\mathbf{q})\chi_{\mathrm{BA}%
}(\mathbf{q})\nonumber \\
&  -\sum_{\mathbf{q}}J_{\mathrm{B}\text{,}\mathbf{R}}\left(  \mathbf{q}%
\right)  J_{\mathrm{B},\mathbf{R}^{\prime}}(-\mathbf{q})\chi_{\mathrm{BB}%
}(\mathbf{q}).
\end{align}
$\chi=\left(
\begin{array}
[c]{cc}%
\chi_{\mathrm{AA}} & \chi_{\mathrm{AB}}\\
\chi_{\mathrm{BA}} & \chi_{\mathrm{BB}}%
\end{array}
\right)  $ is renormalized spin susceptibility from random phase approximation
(RPA) calculation, $\chi=\frac{\chi_{0}}{1-U\chi_{0}}$ where $\chi_{0}\left(
\mathbf{q}\right)  $ is defined by $\chi_{0}\left(  \mathbf{q}\right)
=-\frac{1}{N_{s}}\sum_{\mathbf{k}}G_{\sigma}\left(  \mathbf{k}\right)
G_{-\sigma}\left(  \mathbf{k}-\mathbf{q}\right)  .$

In the continuum limit ($\mathbf{q}\rightarrow0$) and weak interaction case
($U/t<1$), the spin susceptibility in terms of microscopic variables is
$\chi_{0}\sim|\mathbf{q}|.$ So we can easily derive that $J_{\mathrm{R}}%
\sim \int \frac{1}{\left \vert \mathbf{q}\right \vert }e^{i\mathbf{q}%
\cdot \mathbf{R}}d^{2}\mathbf{q}\sim \frac{1}{\left \vert \mathbf{R}\right \vert
}.$ The $\frac{1}{R}$ decay rate of the RKKY coupling between the intrinsic
MIs is much different with $\frac{1}{R^{3}}$ decay rate of the RKKY coupling
for two extrinsic (Anderson) MIs. Fig.(2a) and Fig.(2c) show the numerical
results of $J_{\mathrm{R}}$ that is indeed a function of the impurity distance
$R=|\mathbf{R}-\mathbf{R}^{\prime}|$ along both zigzag direction and armchair
direction. The RKKY coupling is FM for two vacancies on the same sublattice
(we denote the case by \textrm{AA/BB}) as $J_{\mathrm{R}}(\mathbf{R}\in A/B,$
$\mathbf{R}^{\prime}\in A/B)<0$ and AFM for two vacancies on the different
sublattices (we denote the case by \textrm{AB/BA}) as $J_{\mathrm{R}%
}(\mathbf{R}\in A/B,$ $\mathbf{R}^{\prime}\in B/A)>0$ that has been predicted
before \cite{rkky}.

For the weak interacting case, the RKKY interaction is proportion to $U^{2}$
as $J_{\mathrm{R}}\rightarrow U^{2}.$ On the other hand, due to the magnetic
instability near $U\rightarrow U_{c}=2.23t,$ the renormalized spin
susceptibility $\chi$ diverges. Thus, we found that the RKKY coupling also
diverges near the quantum critical point at $U_{c}$, as $J_{\mathrm{R}%
}\rightarrow(U_{c}-U)^{-1}$\cite{black}. See the results in Fig.(2b) and Fig.(2d).

At second step, we study the direct Heisenberg exchange (DHE) coupling. The
DHE coupling between two QLSMs on $\mathbf{R}$ and $\mathbf{R}^{\prime}$ is
described by the following Hamiltonian
\begin{equation}
J_{\mathrm{D}}(\mathbf{R},\mathbf{R}^{\prime})\mathbf{\hat{S}}_{\mathbf{R}%
}\cdot \mathbf{\hat{S}}_{\mathbf{R}^{\prime}}%
\end{equation}
where $J_{\mathrm{D}}(\mathbf{R},\mathbf{R}^{\prime})=-U\sum_{i}\left \vert
\psi_{0,\mathbf{R}_{i}}\right \vert ^{2}\left \vert \psi_{0,\mathbf{R}%
_{i}^{\prime}}\right \vert ^{2}$ is the coupling strength which is always
negative (or $J_{\mathrm{D}}<0$). $\psi_{0,\mathbf{R}_{i}}$ and $\psi
_{0,\mathbf{R}_{i}^{\prime}}$ are wave functions of the two quasi-localized
states of the vacancies at $\mathbf{R}$ and $\mathbf{R}^{\prime}$.

\begin{figure}[ptb]
\includegraphics* [width=0.5\textwidth]{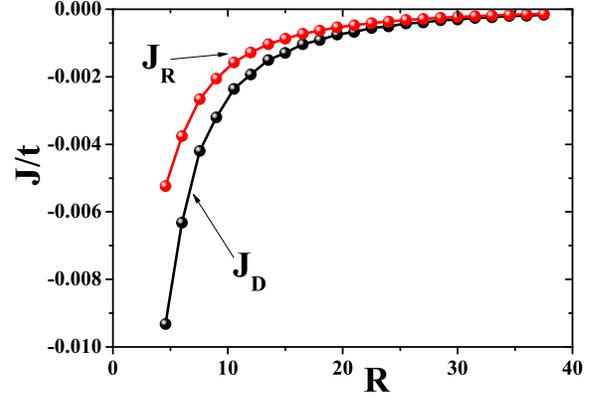}\caption{(Color online) The
DHE coupling between two QLSMs on the same sublattice along armchair direction
for the case of $U=1.5t$. }%
\end{figure}

The wave function of the quasi-localized state exactly distributes only on the
opposite sub-lattices. Therefore, the DHE coupling between two QLSMs around
vacancies on the different sublattices vanishes, $J_{\mathrm{D}}(\mathbf{R}\in
A/B,\mathbf{R}^{\prime}\in B/A)=0$. On the contrary, the DHE coupling between
two QLSMs of vacancies on the same sublattice is finite, $J_{\mathrm{D}%
}(\mathbf{R}\in A/B,\mathbf{R}^{\prime}\in A/B)<0$. In Fig.(3),\ we calculate
the DHE coupling of two defects (see the black line). The fit decay rate from
the numerical calculations for $R<38$ is about $R^{-1.412}$ along zigzag
direction and $R^{-1.644}$ along armchair direction. However, for the WLSMs of
Anderson MIs, the DHE coupling between the localized spin moments can be
definitely ignored.

\begin{figure}[ptb]
\includegraphics* [width=0.5\textwidth]{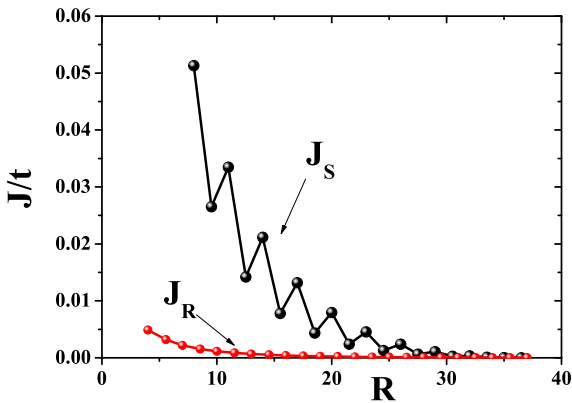}\caption{(Color online) The
SE coupling between two QLSMs on the difference sublattices along armchair
direction for the case of $U=1.5t$.}%
\end{figure}

At third step, we study the superexchange (SE) coupling. When there are two
vacancies on different sublattice nearby, the wave-functions of the zero modes
around different vacancies may overlap and the quantum tunneling effect
occurs. As a result, the energy degeneracy of the localized states is removed
and we may have a finite energy level splitting. Taking the tight-binding
limit, we can regard the quasi-localized states to obtain the sets of wave
functions $\psi_{0,i}(\mathbf{R}),$ where $\sigma=\uparrow,$ $\downarrow$
denote spin degree of freedom and $\mathbf{R}$ denotes the position of the
impurity. The quantum tunneling effect leads to an effective hopping of
electrons from one quasi-localized state to another. The effective model of
the zero modes becomes $t_{RR^{\prime}}\left(  \hat{\alpha}_{\mathbf{R}%
,\sigma}^{\dagger}\hat{\alpha}_{\mathbf{R}^{\prime},\sigma}+h.c.\right)  $
where $\hat{\alpha}_{\mathbf{R},\sigma}$ is an annihilation operator around a
vacancy and the hopping strength $t_{RR^{\prime}}$ is just the energy
splitting $\Delta E_{RR^{\prime}}$ from the quantum tunneling as
$t_{RR^{\prime}}\rightarrow \Delta E_{RR^{\prime}}/2$. After considering the
effective on-site interaction $U_{\mathrm{eff}}$, the effective Hamiltonian of
the electrons on the two quasi-localized states is given by
\begin{align}
H_{\text{\textrm{eff}}}  &  =-t_{RR^{\prime}}(\hat{\alpha}_{\mathbf{R},\sigma
}^{\dagger}\hat{\alpha}_{\mathbf{R}^{\prime},\sigma}+h.c.)\nonumber \\
&  -\mu_{\mathrm{eff}}\sum_{\mathbf{R}}n_{\mathbf{R},\sigma}+U_{\mathrm{eff}%
}\sum_{\mathbf{R}}\hat{n}_{\mathbf{R}\uparrow}\hat{n}_{\mathbf{R}\downarrow}.
\end{align}

At half filling $\mu_{\mathrm{eff}}=0,$ for the two QLSMs, there exists an
energy splitting between the singlet state and the triplet states. We get an
effective SE term $J_{\text{\textrm{S}}}(\mathbf{R},\mathbf{R}^{\prime
})\mathbf{\hat{S}}_{\mathbf{R}}\cdot \mathbf{\hat{S}}_{\mathbf{R}^{\prime}}$
where $J_{\text{\textrm{S}}}(\mathbf{R},\mathbf{R}^{\prime})$ is the SE
coupling strength. For two vacancies on the same sublattice, $J_{\mathrm{S}%
}(\mathbf{R}\in A/B,$ $\mathbf{R}^{\prime}\in A/B)=0$. For two vacancies on
the different sublattices, $J_{\text{\textrm{S}}}(\mathbf{R}\in A/B,$
$\mathbf{R}^{\prime}\in B/A)>0$. In the strong coupling limit,
$U_{\mathrm{eff}}/t_{RR^{\prime}}\rightarrow \infty,$ $J_{\text{\textrm{S}}%
}\simeq \frac{4\left \vert t_{RR^{\prime}}\right \vert ^{2}}{U_{\mathrm{eff}}}$.
Due to $\left \vert t_{RR^{\prime}}\right \vert \rightarrow \frac{1}{R}$ for
$R\rightarrow \infty,$ we have $J_{\text{\textrm{S}}}(\mathbf{R},\mathbf{R}%
^{\prime})\simeq \frac{4\left \vert t_{RR^{\prime}}\right \vert ^{2}%
}{U_{\mathrm{eff}}}\rightarrow \frac{1}{R^{2}}$. For the weak coupling case, we
calculate the SE coupling strength numerically. The results are given in
Fig.(4). However, for the WLSMs of Anderson MIs, there is no the SE coupling
between the localized spin moments.

\begin{widetext}
\begin{table*}[t]%
\begin{tabular}
[c]{|c|c|c|} \hline & Vacancy-induced magnetic impurity in graphene & Anderson magnetic impurity in graphene \\
\hline Type of magnetic impurity& Intrinsic & Extrinsic \\
\hline Spin moment & Quasi-localized & Well-localized \\
\hline s-d hybridization & $0$ & Finite \\
\hline s-d coupling & Non-local FM Hund rule's coupling & Local AFM Kondo coupling \\
\hline Kondo effect  & No screening effect (?) & Kondo effect in pseudo-gap system\\
\hline Decay rate of RKKY coupling  & $R^{-1}$ & $R^{-3}$ \\
\hline Decay rate of DHE coupling  & $R^{-\alpha}$ for AA/BB case; $0$ for AB/BA case& $0$ \\
\hline Decay rate of SE coupling  & $R^{-2}$ for AB/BA case; $0$ for AA/BB case& $0$ \\
\hline
\end{tabular}
\caption{The differences between intrinsic MI and extrinsic MI in graphene. '?' means that we are not sure about this results from mean field calculations. $\alpha$ is about $1.412$ along zigzag direction and $1.644$ along armchair direction.}
\end{table*}
\end{widetext}

In summary, we derive the effective coupling between two QLSMs. For
\textrm{AA/BB} case, the total coupling between two QLSMs is
$J_{\text{\textrm{D}}}+J_{\mathrm{R}}.$ Now both $J_{\text{\textrm{D}}}$ and
$J_{\mathrm{R}}$ are negative. So we have an FM coupling. We also compare
$\left \vert J_{\text{\textrm{D}}}\right \vert $ and $\left \vert J_{\mathrm{R}%
}\right \vert $ and find that $\left \vert J_{\text{\textrm{D}}}\right \vert
>\left \vert J_{\mathrm{R}}\right \vert $ for short distance between two
vacancies while $\left \vert J_{\text{\textrm{D}}}\right \vert <\left \vert
J_{\mathrm{R}}\right \vert $ for large distance. The critical distance
dependents on the on-site interaction $U$. In Ref.\cite{pis}, a decay rate of
$R^{-1.43}$ ($R^{-1.644}$) for moments along zigzag (armchair) direction
separations of up to $25$\textrm{\AA }\ had been extracted. Our results
($R^{-1.412}$) match their calculations. For \textrm{AB/BA} case, the coupling
between two QLSMs is $J_{\text{\textrm{S}}}+J_{\mathrm{R}}.$ Now both
$J_{\text{\textrm{S}}}$ and $J_{\mathrm{R}}$ are positive. So we have an AFM
coupling. We also compare $J_{\text{\textrm{S}}}$ and $J_{\mathrm{R}}$ and
find that $J_{\text{\textrm{S}}}>J_{\mathrm{R}}$ for short distance between
two vacancies while $J_{\text{\textrm{S}}}<J_{\mathrm{R}}$ for large distance.
Thus, for the case of graphene with $U=1.5t$, from the results in Fig.(3) and
Fig.(4), the RKKY coupling is always smaller than SE coupling or DHE coupling
when the distance between two vacancies is short, $R<38$. These results that
seem to contradict to people's intuition can be naturally understood from the
quasi-localization of the intrinsic MI induced by vacancies in graphene. That
means people had over-estimated the contribution of RKKY coupling during
studying the quantum magnetism of graphene with vacancies.

In the end we draw the conclusions. In this paper we developed a theory for
the intrinsic MIs with the QLSM induced by the vacancies in graphene. Because
the intrinsic MIs are characterized by the zero modes that {are orthotropic to
the itinerant electrons, }their properties are much different to those of
traditional Anderson MIs with the WLSMs. We give a table to compare the
principal features of the two types (intrinsic and extrinsic) of MIs in
graphene. Furthermore, the theory for the intrinsic MIs with the QLSM induced
by the vacancies can be generalized to other bipartite system with
particle-hole symmetry.

\begin{center}
{\textbf{* * *}}
\end{center}

This work is supported by National Basic Research Program of China (973
Program) under the grant No. 2011CB921803, 2012CB921704 and NSFC Grant No.
11174035.


\bigskip

\begin{thebibliography}{99}                                                                                               %


\bibitem {nov1}K. S. Novoselov, et.al, Science, \textbf{306}, 666 (2004).

\bibitem {nov2}K. S. Novoselov, et.al, Nature, \textbf{438}, 197 (2005).

\bibitem {neto}C. Neto A H, et.al, Rev. Mod. Phys. \textbf{81}, 109 (2009).

\bibitem {sarma}S Das Sarma, et al, Rev. Mod. Phys. \textbf{83}, 407 (2011).

\bibitem {aw}S. A. Awschalom, et.al, Science, \textbf{294},~1488 (2001).

\bibitem {zero-mode}V. M. Pereira, et.al, Phys. Rev. Lett. \textbf{96,} 036801 (2006).

\bibitem {ya}O. V. Yazyev, L. Helm, Phys. Rev. \textbf{B 75}, 125408 (2007).

\bibitem {zero1}F. Ducastelle, Phys. Rev. \textbf{B 88}, 075413 (2013).

\bibitem {nair12}R. R. Nair, et.al, Nature Phys. \textbf{8,} 199 (2012).

\bibitem {hong12}Hong X, et.al, Phys. Rev. Lett. \textbf{108} 226602 (2012).

\bibitem {rkky}M. A. H. Vozmediano, et al, Phys. Rev. \textbf{B 72}, 155121
(2005); V. V. Cheianov and V. I. Fal'ko, Phys. Rev. Lett., \textbf{97},~226801
(2006); V. K. Dugaev, et al, Phys. Rev. \textbf{B 74}, 224438 (2006); V. K.
Dugaev, et al, Phys. Rev. \textbf{B 74},~224438 (2006); S. Saremi, Phys. Rev.
\textbf{B 76}, 184430 (2007); L. Brey et al, Phys. Rev. Lett. \textbf{99,}
116802 (2007); E. Hwang, et al, Phys. Rev. Lett. \textbf{101}, 156802 (2008);
J. E. Bunder et al, Phys. Rev. \textbf{B 80}, 153414 (2009); P. Venezuela, et
al, Phys. Rev.\textbf{ B 80}, 241413(R) (2009); A. M. Black-Schaffer, Phys.
Rev. \textbf{B 81}, 205416 (2010); M. Sherafati, et al, Phys. Rev. \textbf{B
83}, 165425 (2011); S. R. Power, et al, Phys. Rev. \textbf{B 83}, 155432
(2011); B. Uchoa, et al, Phys. Rev. Lett. \textbf{106}, 016801; (2011); H.
Lee, et al, Phys. Rev. \textbf{B 85}, 075420 (2012). S. R. Power, M. S.
Ferreira, Crystals, \textbf{3 }(1), 49 (2013).

\bibitem {kondo}B. Uchoa, et al, Phys. Rev. Lett. \textbf{106}, 016801 (2011).
M. A. Cazalilla, et al, arXiv:1207.3135.

\bibitem {kondo1}L. Fritz, M. Vojta, Rep. Prog. Phys. \textbf{76}, 032501 (2013).

\bibitem {fuhrer11}J. H. Chen, et.al, Nature Phys. \textbf{7,} 535 (2011).

\bibitem {comment}For the case with a little next nearest neighbor hopping,
the wave-function around a lattice-defect slightly permeates onto A
sublattice. And the energy of the localized states shifts from zero to a
finite value.

\bibitem {black}Black-Schaffer, Phys. Rev. \textbf{B 82},~073409 (2010).

\bibitem {pis}L. Pisani, et al, New J. Phys. \textbf{10},~033002 (2008).
\end{thebibliography}
\end{document}